\documentclass[%
 reprint,
 amsmath,amssymb,
 aps,superscriptaddress
]{revtex4-1}

\usepackage{graphicx}
\usepackage{dcolumn}
\usepackage{bm}
\usepackage[mathscr]{euscript}
\usepackage{mathrsfs}
\usepackage{amsbsy}
\usepackage{color}
\usepackage{scrextend, titlesec, braket, emptypage, setspace, etoolbox,  hyperref}

\newcommand{\vkap}{\varkappa}
\newcommand{\vphi}{\varphi}

\begin{document}


\title{Axion-like particle production in a laser-induced dynamical spacetime}

\author{M. A. Wadud}
\affiliation{Department of Physics, University of Oxford, Parks Road, Oxford OX1 3PU, UK}
\affiliation{School of Physics \& Astronomy, 116 Church Street S.E., Minneapolis, MN 55455, USA}
\author{B. King}
\affiliation{School of Computing, Electronics and Mathematics, Plymouth University, Devon, PL4 8AA, UK}
\author{R. Bingham}
\affiliation{University of Strathclyde, Glasgow, G4 0NG, UK}
\affiliation{Rutherford Appleton Laboratory, Chilton, Didcot OX11 0QX, UK}
\author{G. Gregori}
\affiliation{Department of Physics, University of Oxford, Parks Road, Oxford OX1 3PU, UK}
%


\date{\today}

\begin{abstract}
We consider the dynamics of a charged particle (e.g., an electron) oscillating in a laser field in flat spacetime and describe it in terms of the variable mass metric.
By applying Einstein's equivalence principle, we show that, after representing the electron motion in a time-dependent manner, the variable mass metric takes the form of the 
Friedmann-Lema\^{i}tre-Robertson-Walker metric. We quantize a massive complex scalar or pseudo-scalar field in this spacetime and derive the production rate of electrically neutral, spinless particle-antiparticle pairs. We show that this approach can provide an alternative experimental method to axion searches. 
\end{abstract}

\maketitle


It is well known that particle-production phenomena can occur in a curved or dynamic spacetime \cite{Birrell:1982q}. For example, thermal radiation can arise from particle production near the event horizon of a black hole, an effect commonly known as the Hawking radiation \cite{r7,r8}. This is a quite general fact, not confined to black holes. As hypothesized by Davies, Unruh and Fulling \cite{r9,r10,r11}, an observer in a uniformly accelerated frame experiences the surrounding vacuum as filled with thermal radiation with temperature
$T_{DU} = \hbar a/2\pi k_B c = 4.05 \times 10^{-23} a$ K, where $a$ is the acceleration (in cm/s$^2$) and $k_B$ is the Boltzmann constant.
The expansion of the universe also gives rise to a curved metric called the Friedmann-Lema\^{i}tre-Robertson-Walker (FLRW) metric: $ds^2=dt^2-h^2(t) d\mathbf{x}^2$. Here $h(t)$ is the scale factor which quantifies the relative expansion of the universe. In the FLRW metric, particles are spontaneously produced as a result of the expansion of the universe \cite{Parker:1968pc, Parker:1969qf,Mensky:1980ki, Parker:2012at}. Of particular interest is the inflationary period, from $10^{-36}\,\mathrm{s}\mathrm{\:until\:}10^{-32}\,\mathrm{s}$ after the big bang. During this time, 
it is thought that
the universe expanded exponentially, and spacetime was highly curved and dynamic. Understanding particle production during inflation \cite{Campos:1991ff, Zeldovich1974c, Zeldovich:1971mw} may help answer major questions like why the universe today is isotropic and flat, and why there is more matter than antimatter \cite{linder1990particle, Lyth:1998xn}.

The latter is an example of a spontaneously broken symmetry that may require the existence of particles beyond the standard model. The axion is one of such particles, a pseudo Nambu-Goldstone boson which arises from the spontaneously broken Peccei-Quinn symmetry \cite{peccei}. 
Both astrophysical bounds from stars and galaxies \cite{Raffelt1986,Conlon2013} as well as laboratory searches \cite{Graham2015,Rosenberg2015a} have provided limits for the mass and coupling constants of these hypothetical particles. While experimental searches so far have not yet identified an axion candidate, the parameter space left to explore is still large and there is a need of more sensitive probes before the axion existence can be confidently ruled out.

Recent advancements in ultra-high intensity lasers \cite{Strickland:1985c} have stirred interest in the possibility of detecting both the Schwinger effect and dynamic spacetime phenomena \cite{Mourou:1998u,Bulanov:2003zz,Ahlers2008,Piazza:2012e}. Projects under development include the European Extreme Light Infrastructure \cite{eli}, which will provide radiation beams of intensities exceeding $10^{24}\,\mathrm{\:W/cm^{2}}$; the X-ray free electron lasers (XFEL) based at DESY Hamburg, and the LCLS (Linac Coherent Light Source) facility at SLAC, where highly tunable x-ray pulses with narrow bandwidth and high intensity are already available. Here we propose a mechanism for electrically neutral particle-antiparticle pair production in a laser field, whereby the variation of the metric around a charged particle oscillating in the laser field gives rise to spontaneous particle-antiparticle pair production.

We start with the Lagrangian density of a free, massive, minimally-coupled real scalar or pseudo-scalar field $\phi(x)$ under the FLRW metric\ $g_{\mu\nu}=h^2(\eta)\eta_{\mu\nu}$, where $\eta_{\mu\nu}=\mathrm{diag}(1,-1,-1,-1)$ is the Minkowski metric and $h(\eta)$ is the scale factor \cite{Mukhanov2007i}:
\begin{equation}
\mathscr{L}=\frac{1}{2}\sqrt{-\det g_{\mu\nu}} \left[g^{\mu\nu} (\partial_{\mu}\phi)( \partial_{\nu}\phi)-m^2 \phi^2\right].
\end{equation}
Here $x^0=\eta$ is the conformal time (not to be confused with the Minkowski metric $\eta_{\mu\nu}$) which is related to the physical time $t$ by $dt/d\eta=h(\eta)$.
Natural units $(\hbar=c=\epsilon_0=1)$ are used in all the derivations, and conversion to physical units will be explicitly mentioned.

The equation of motion of the field is the one that extremises the action functional $S=\int d^4x \mathscr{L}$. The extremisation condition is equivalent to Euler-Lagrange equation, giving for our case the Klein-Gordon equation. The next step consists in the procedure of canonical quantisation of the field $\phi$ and provide a framework for particles to be created and annihilated.
In FLRW spacetime, however, the vacuum states at different times, $\ket{0}_{\eta_0}$ and $\ket{0}_{\eta_1}$ are different, and a notion of particle and antiparticle that is consistent at all times is unattainable. 
To circumvent the ambiguity about the vacuum state, we first assume the existence of a preferred particle model that provides time-independent creation/annihilation operators  from which we can construct a reference vacuum state. Such conditions are fulfilled, for example, when looking at the solution of the Klein-Gordon equation at 
asymptotic times $(\eta \to \pm \infty)$ \cite{Fulling:1974pu,Kluger:1998bm,Calzetta:2008n}. 
These are used to define time-dependent creation and annihilation operators, related to the asymptotic ones by Bogoliubov transformation.
The procedure outlined above corresponds to the kinetic approach to quantum field theory, leading to the so called quantum Vlasov equation \cite{Kluger:1998bm,Hebenstreit2008}. We obtain (see {\it e.g.}, \cite{Calzetta:2008n}),
\begin{widetext}
\begin{equation}
\begin{split}
\frac{d\mathscr{N}_{\mathbf{k}}}{d\eta}&(\eta)=\frac{\dot\omega_k(\eta)}{2\omega_k(\eta)}\int_{\eta_0}^{\eta}d\eta'\frac{\dot\omega_k(\eta')}{\omega_k(\eta')} [1+2\mathscr{N}_{\mathbf{k}}(\eta')]\cos[2\Theta_k(\eta)-2\Theta_k(\eta')],
\label{eq:vlasov}
\end{split}
\end{equation}
\end{widetext}
where $\mathscr{N}_{\mathbf{k}}$ is the time-dependent number of pairs of spatial mode
$\mathbf{k}$,
\begin{equation}
\omega_k^2=\mathbf{k}^2+m^2h^2-\frac{\ddot h}{h},
\label{eq:freq}
\end{equation}
with the dot notation representing differentiation by $\eta$ (i.e. $\ddot{h}=d^{2}h/d\eta^{2}$) and
\begin{equation}
\Theta_k(\eta)=\int^{\eta}d\eta'\: \omega_k(\eta').
\label{eq:theta}
\end{equation}
The time $\eta_0$ is defined such that $\mathscr{N}_{\mathbf{k}}(\eta_0)=0$.
The quantum Vlasov equation, eq. (\ref{eq:vlasov}), is formally similar to the one obtained by Kluger et al. \cite{Kluger:1998bm} for bosonic pair production in flat spacetime under an oscillating electron field. However, in our case, it has been specialized such that
there is no explicit presence of an electric field and the spacetime is more generally defined by the FLRW metric. 
This is in fact generally the case for field theories in background fields. Since the particle number operator does not, in general, commute with the interaction Hamiltonian, one must be cautious interpreting results at intermediate times. Different particle number definitions that coincide at asymptotic times, may disagree by orders of magnitude at intermediate times (this phenomenon has been recently studied using the superadiabatic basis to analyse the Schwinger effect \cite{dunne14c}). We note the quantum Vlasov equation's non-Markovian character  \cite{Schmidt:1998zh}: the term $1+2\mathscr{N}_{\mathbf{k}}(\eta')$ in the integral means that the equation is non-local in time, i.e., the production rate of pairs is dependent on the history of the system.

Having obtained the particle production rate in an expanding spacetime, we now describe the dynamics of a particle in a laser field with an alternative metric that, as we shall see, bears many resemblances with the FLRW metric. In our approach we do not quantise the laser electromagnetic field nor the metric, which will be treated classically as existing in the background. Following closely the derivation by Crowley et al. \cite{Crowley:2012t,Gregori}, we consider the dynamics of a free particle of mass $m_0$ under the variable mass metric \cite{Narlikar1993}. The name of the metric derives from the appearance of the ``variable mass" $h m_0$ in the place of the rest mass $m_0$ in dynamical equations that are similar to flat spacetime equations. We have $g_{\mu\nu}=h^2(\mathbf{x})\eta_{\mu\nu}$, where $\eta_{\mu\nu}$ is the Minkowski metric and $h(\mathbf{x})$ is a spatial field. In general relativity, the dynamics of a free particle of mass $m_0$ in a spacetime with metric $g_{\mu \nu}$ is determined by its Lagrangian \cite{Poisson:2014g} $L=-(g_{\mu \nu} v^{\mu} v^{\nu})^{1/2}m_0$, where $v^{\mu}=dx^{\mu}/dx^{0}$ is the 4-velocity. The Lagrangian for the variable mass metric is thus \cite{Crowley:2012t}, $L=-h m_0/\gamma$, where $\gamma=(1-v^2)^{-1/2}$. The canonical 3-momentum is given by
$\mathbf{p}=\partial L/\partial \mathbf{v}=\gamma h m_0 \mathbf{v}$, and
the Hamiltonian is then $H=\mathbf{p}\cdot \mathbf{v}-L=\gamma h m_0$. 
Using Hamilton's equations then obtain:
\begin{equation}
H=(\mathbf{p}^2+h^2m_0^2)^{1/2},
\label{eq:hamilton2}
\end{equation}
and,
\begin{equation}
\mathbf{a}=\mathbf{\dot{v}}=-\frac{1}{\gamma^2}\frac{\partial \ln h}{\partial \mathbf{x}}
\label{eq:acc1}
\end{equation}
where we have used the fact that the Hamiltonian has no explicit time dependence.

Let us now consider the dynamics of a charged particle, with charge $q$ and mass $m_0$, oscillating, with frequency $\nu$, in a laser field with vector potential $\mathbf{A}$ and null scalar potential (Coulomb radiation gauge) in flat spacetime, $\eta_{\mu\nu}$. The Lagrangian for this particle is
$L =- m_0/\gamma + q \mathbf{v}\cdot\mathbf{A}$, which gives us the canonical momentum
$\mathbf{p}=\gamma m_0 \mathbf{v}+q\mathbf{A}$ and the Hamiltonian
$H=[(\mathbf{p}-q\mathbf{A})^2+m_0^2]^{1/2}$. 
We can decompose the momentum into parallel and perpendicular components with respect to $\bf A$, that is $\mathbf{p} = \mathbf{p_\parallel}+
\mathbf{p_\perp}$. Thus if $\mathbf{v}_0$ is the velocity of the particle due to the influence of the laser field, and any remaining components are sufficiently small, we approximately have 
$(\mathbf{p_\parallel}-q\mathbf{A})^2 \approx \gamma_0^2 m_0^2 v_0^2$, with $\gamma_0=(1-v_0^2)^{-1/2}$. 
We notice that $|\mathbf{p}| \sim |\mathbf{p_\perp}| \approx \gamma_1 \gamma_0 m_0 v_1$, where $\gamma_1=(1-v_1^2)^{-1/2}$ is calculated with respect to a particle velocity $v_1$ which is not associated to the motions induced by the laser field (that is, $\mathbf{v}_1 = \mathbf{v}-\mathbf{v}_0$, and $\mathbf{v}_1 \perp \mathbf{v}_0$). This holds under the condition that either $v_0 \ll 1$ or $v_1 \ll 1$.
Hence,
\begin{equation}
H=(\mathbf{p}^2+\gamma_0^2 m_0^2)^{1/2},
\label{eq:hamilton3}
\end{equation}
and
\begin{equation}
\mathbf{a}=-\frac{1}{\gamma_1^2}\frac{\partial \ln \gamma_0}{\partial \mathbf{x}}.
\label{eq:acc3}
\end{equation}
We immediately notice that, if we make the substitutions 
$\gamma_0 \rightarrow h$ and $\gamma_1 \rightarrow \gamma$, these last two equations are the same as (\ref{eq:hamilton2}) and (\ref{eq:acc1}). Einstein's equivalence principle asserts that an observer cannot distinguish between his frame's acceleration in flat spacetime and a metric field 
whose geodesic has equal acceleration, {\it i.e.}, the physics is the same in both cases. Hence, the dynamics of the charged particle oscillating in the laser field in flat spacetime may be equivalently described by the variable mass metric Hamiltonian of a free particle.

The idea that electromagnetic acceleration can give rise to dynamics that can be equivalently described by the variable mass metric will now be used to represent $h$ in a time-dependent functional form so that it becomes equivalent to the scale factor of the FLRW metric. This key result will allow us to employ the field quantisation formalism developed earlier and obtain the particle-antiparticle pair production rate with the quantum FLRW-Vlasov equation. In the frame of the charged particle the vacuum acquires a finite number of particles. However, this change has an observable signature in the laboratory frame only if the vacuum in the accelerated frame couples with a detector \cite{Unruh:1976db}. To accomplish this, we assume that the accelerated motion occurs in the presence of an external magnetic field, ${\bf B}_{\rm ext}$, aligned with the velocity of the charge. Under these conditions, the Lagrangian density is modified by an extra term which describes the coupling of the axion field with the photons, given by \cite{Raffelt1986}
\begin{equation}
\mathscr{L}_a=\sqrt{-\det g_{\mu\nu}} \frac{1}{M} {\pmb{\mathscr{E}}} \cdot{{\bf B}_{\rm ext}} \phi,
\end{equation}
where $\pmb{\mathscr{E}}$ describes the electric field of the photons, and $1/M \equiv \alpha g_\gamma / \pi f_a$ is the coupling constant, with $\alpha$ the fine structure constant, $g_\gamma$ a coefficient of order unity which depends on the 
details of the axion model, and $f_a$ the axion decay constant \cite{Raffelt1986}. That is, in the accelerated frame the axions forming the vacuum couples with the external magnetic field producing photons. This is an observable signature.
If we assume that the photon and the axion fields propagate with the same direction and phase, then the additional term in the Lagrangian density leads to a modified dispersion relation (\ref{eq:freq}),
\begin{equation}
\omega_k^2=\mathbf{k}^2+\left(m^2+\frac{B_{\rm ext}^2}{2 M^2}\right) h^2-\frac{\ddot h}{h}.
\label{eq:freq2}
\end{equation}
We notice that only the axions that interact with the external magnetic fields are the ones that are observed in the laboratory frame. The external magnetic field is the same both in the laboratory and accelerated frames. In the following, we have set ${\bar m}^2=m^2 \left(1+B_{\rm ext}^2/2 M^2 m^2\right)$. 

We now describe the acceleration of a charge particle on mass $m_0$ in a strong laser field $\bf E$, and in presence of a much weaker, constant, external magnetic field ${\bf B}_{\rm ext}$ (see above). We thus assume that the motion of the charged particles is determined by the laser field only. We take the laser pulse of frequency $\nu$, four-wavevector $\vkap$, phase $\varphi=\vkap\cdot x$, duration $\tau$ and intensity parameter $\xi=qE/m_{0}\nu$ to be represented at the focus by a vector potential
\begin{equation}
\mathbf{A} = ~m_{0}\xi \exp\left[-\left(\frac{\varphi}{\Phi}\right)^{2}\right]~\cos\varphi ~ \mathbf{\hat{z}},
\end{equation}
where $\Phi=\nu \tau$ and $\mathbf{\hat{z}}$ is the unit vector in the $z$-direction. We limit the analysis to non-relativistic electron motion
 ($\gamma_0 \approx 1$), by specifying that $\xi \ll 1$. Assuming the particle begins at the origin with zero momentum in the infinite past, the velocity component in the field direction is $\mathbf{\dot{x}} \cdot \mathbf{\hat{z}} = q A / m_{0}$, which gives:
\[
 h = \left(1-v_{0}^{2}\right)^{-1/2} = \left[1-(qA/m_{0})^{2}\right]^{-1/2}  \approx 1 
+\frac{q^{2}A^{2}}{2m^{2}_{0}}.
\]
We see that $h \geq 1$, meaning that space expands when the electric field is non-zero. 
This can be interpreted as the result of the
increased energy density of free space due to the presence of an electric field. In other terms, the electron acquires an effective mass $m_{\rm eff} = h m_0$. The idea of an effective mass to describe the motion of electrons in intense laser beams is not new, and it is
associated with the frequency shift of the radiation emitted by a particle in an intense electromagnetic field \cite{kibble}.

With this time-dependent form of $h$, the variable mass metric becomes equivalent to the FLRW metric. We can thus use the quantum field formalism developed earlier to estimate the particle production by integrating the FLRW quantum Vlasov equation (\ref{eq:vlasov}). 
In doing so, we note that the field mass $m$ appearing in equation $(\ref{eq:freq2})$ for the frequency $\omega_k$ is not necessarily the same as the mass $m_0$ of the oscillating particle. We assume that the test particle being accelerated by the laser field has mass $m_0$ and charge $q$  while the particles that are being produced as a result of the transformed metric have mass $m$ and no charge.

Next, in the low density regime, that is, when the laser electric field is much smaller than the Schwinger's critical field,
we make the  approximation \cite{Gregori:2010uf}
\begin{equation}
\begin{split}
\mathscr{N}_{\mathbf{k}}(\eta_{0},\eta) \approx &\frac{1}{2}\int_{\eta_0}^{\eta}\int_{\eta_0}^{{\eta}''}  d\eta'' d\eta'\frac{\dot\omega_k(\eta')}{\omega_k(\eta')}\frac{\dot\omega_k(\eta'')}{\omega_k(\eta'')} \times \nonumber \\
&\cos[2\Theta_k(\eta')-2\Theta_k(\eta'')],
\end{split}
\end{equation}
where we have assumed that $\mathscr{N}_{\mathbf{k}}(\eta_0)=0$. The integrand is symmetric with respect to the exchange $\eta' \leftrightarrow \eta''$, which means it is symmetric about the line $\eta'=\eta''$. Hence \cite{Prozorkevich2004t}
\begin{equation}
\begin{split}
\mathscr{N}_{\mathbf{k}}(\eta_{0},\eta) \approx
&\frac{1}{4}\bigg|\int_{\eta_0}^{\eta} d\eta'\frac{\dot\omega_k(\eta')}{\omega_k(\eta')}\exp[2i\Theta_k(\eta')]\bigg|^2,
\label{eq:vlasov3}
\end{split}
\end{equation}
where we have used the fact that the antisymmetry of the factor $\sin \{2\Theta_k(\eta')-2\Theta_k(\eta'')\}$ with respect to the exchange $\eta' \leftrightarrow \eta''$ has null contribution to the integral \cite{Gregori:2010uf, Prozorkevich2004t}.

From (\ref{eq:freq2}) we then have:
\begin{equation}
\begin{split}
\omega_k \dot{\omega}_k=&\bar{m}^2 h {\dot h} -\frac{h {\dddot h}-{\dot h} {\ddot h}}{2h^2}
\end{split} \label{eq:omkomkd}
\end{equation}
where $\omega^2_k(\eta) \approx {\bf k}^2 +\bar{m}^2$. The change in the metric perturbation depends only on the external field, so we have $h=h(\varphi)$, but on the other hand wish to integrate over the conformal time, $\eta$. In general, the dependency $\eta(\vphi)$ can be complicated, but for a plane-wave background we can write $d/d\eta = \Omega^{-1} d/d\vphi$ where $\Omega = \vkap \cdot p/m_0$ is the particle energy parameter \cite{heinzl16}. As previously mentioned, of experimental relevance are the asymptotic values of observables for times long after the laser pulse has passed through the seed electrons \cite{dunne14c}. For this reason, we integrate to finite phases, and in the final calculated observables, take the asymptotic limit. In this vein, $\mathscr{N}_{\mathbf{k}}(-\Omega R,\Omega R) \approx \frac{1}{4} |I_{\mathbf{k}}(R)|^{2}$ where:
\begin{equation}
\begin{split}
 I_{k}(R) = \frac{1}{\omega_{k}^{2}\Omega}
\int_{-R}^{R} \left[\bar{m}^{2}h\dot{h} - \frac{h\,\dddot{h} - \dot{h}\ddot{h}}{2h^{2}}\right] \mbox{e}^{2i\Theta_{k}}\,d\vphi.
\end{split}
\end{equation}

\begin{figure}
\noindent\centering
\includegraphics[draft=false,width=0.48\linewidth]{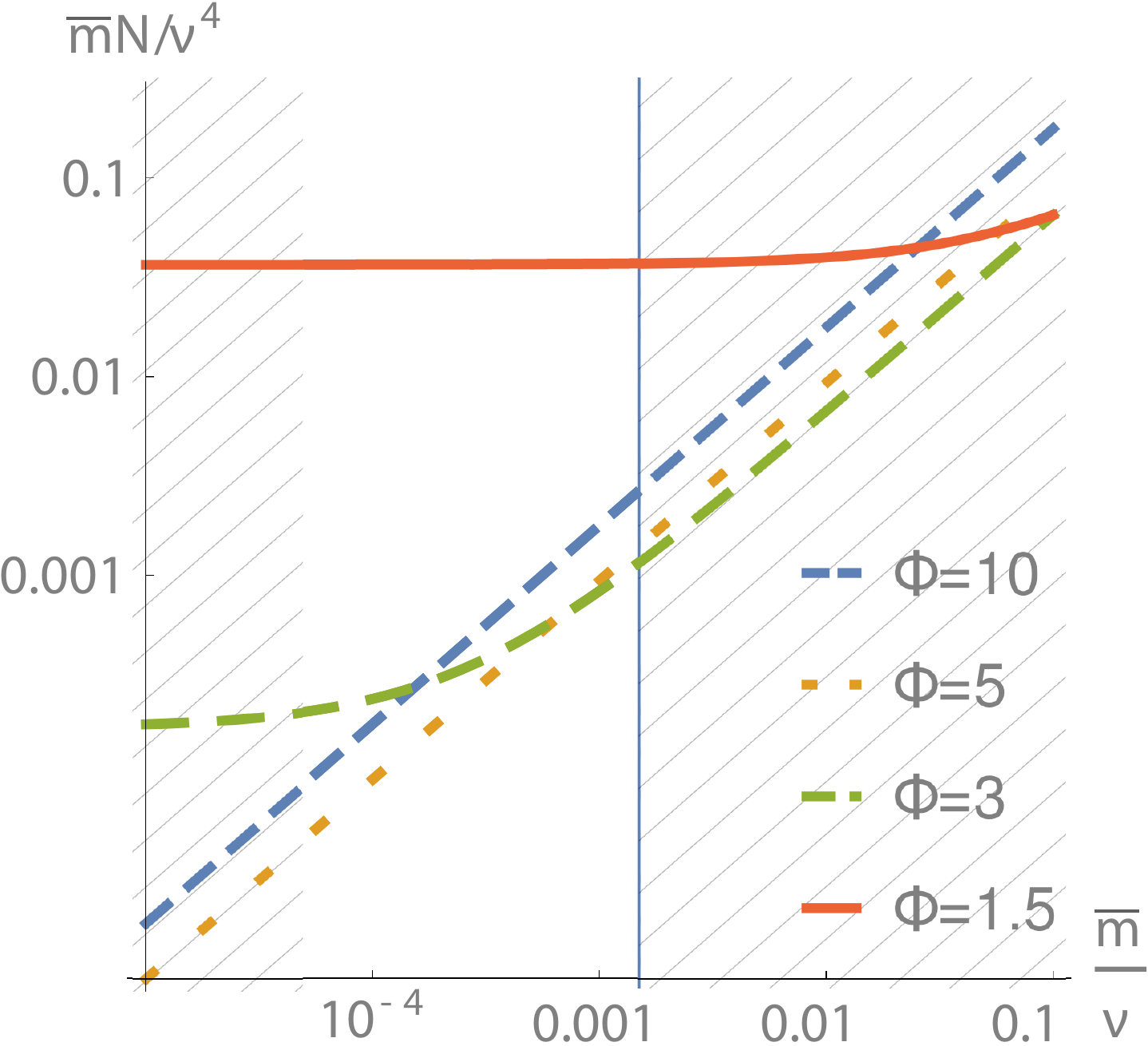}\hfill
\includegraphics[draft=false,width=0.48\linewidth]{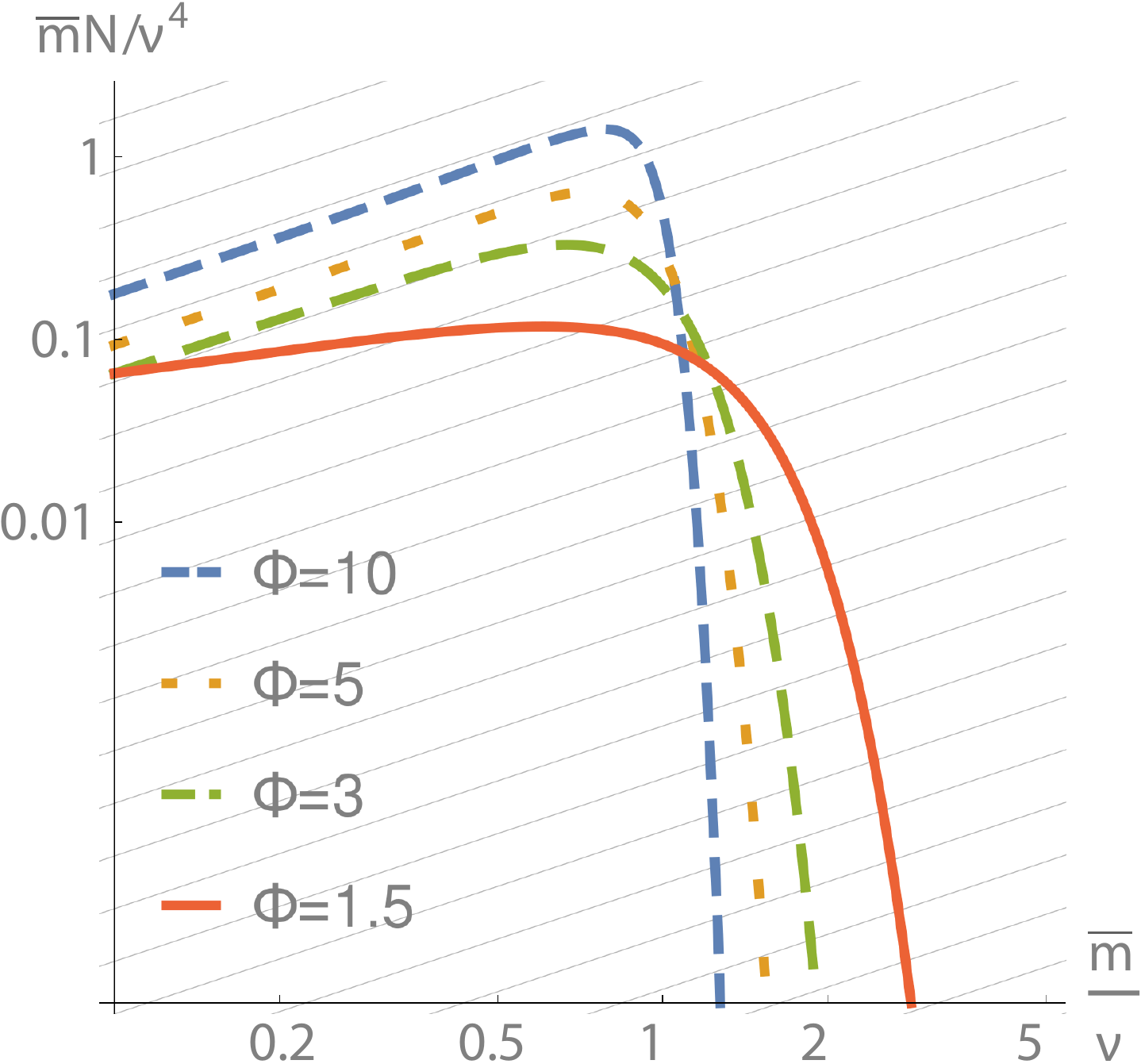}
\caption{Mass dependency of $\bar{m} N/\nu^{4}$ for $\xi=0.1$. Shaded regions indicate axion masses ruled out by the calculation in \cite{borsanyi16}. Left: as $\Phi$ is reduced, $I_{3}$ becomes flat, indicating a dependency $1/\bar{m}$. Right: As $\Phi$ is increased, the heaviest axion mass to which the method is sensitive, decreases.}
\label{fig:plot1} 
\end{figure}

Let us then define $\mathscr{N}_{\mathbf{k}} = \lim_{R\to \infty}\mathscr{N}_{\mathbf{k}}(-\Omega R,\Omega R)$. By assuming the hierarchy $m_{0} \gg m_{0}\xi \gg \Omega \geq \bar{m}$ and expanding to lowest order in $\Phi^{-1}$, terms in pre-exponents of order $O(\bar{m}^{2}\xi^{4}\Omega)$ and $O(\xi^{4}\Omega^{3})$, $O(\bar{m}^{2}\xi^{2}\Omega/\Phi^{2})$ were neglected in the integration. The leading-order terms were then:
\[
 \mathscr{N}_{\mathbf{k}} \approx \frac{\pi\xi^{4}\Phi^{2} (\bar{m}^{2}+2\,\Omega^{2})^{2}}{2^{7}\omega_{k}^{4}}\left[\mbox{e}^{-\frac{\Phi^{2}}{\Omega^{2}}\left(\omega_{k}-\Omega\right)^{2}}\!\!+\mbox{e}^{-\frac{\Phi^{2}}{\Omega^{2}}\left(\omega_{k}+\Omega\right)^{2}}\right]
\]
Integrating over all modes $\mathbf{k}$ gives the total particle density:
\begin{equation}
\begin{split}
N = & \int d^3 \mathbf{k} \,\mathscr{N}_{\mathbf{k}}(\eta) = \frac{\pi\,\Phi^{2}\xi^{4}}{2^{7}} \frac{(\bar{m}^{2}+2\,\nu^{2})^{2}}{\bar{m}} I_{2}(\bar{m}, \nu, \tau),
\end{split}
\end{equation}
\[
I_{2}(\bar{m}, \nu, \tau) = \int_{0}^{\infty} \!\!
\frac{dy~y^{2}}{(y^{2}+1)^{2}} \left[
\mbox{e}^{-\tau^{2}(\omega_{k} - \nu)^{2}} +
\mbox{e}^{-\tau^{2}(\omega_{k} + \nu)^{2}}\right],
\]
(recalling $\Phi = \nu \tau$), $\omega_{k}^{2} = \bar{m}^{2}(1+y^{2})$ and we have assumed the particle starts at rest, so $\Omega = \nu$. Using the Laplace method for the case $\Phi \gg 1$ and a perturbative expansion for when $\bar{m}\tau < 1$, we find that if $\bar{m} \ll \nu$:
\begin{equation}
N \approx
     \dfrac{\pi^{2}\,\xi^{4}}{2^{7}} \dfrac{\nu^{4}}{\bar{m}}~\Phi^{2}\,\mbox{e}^{-\Phi^{2}}. 
     \label{eqn:cases} 
\end{equation}
In the short-pulse case, the produced axion density tends to the familiar $N \sim 1/\bar{m}$. The dependency of axion production on the axion mass is given in Fig. \ref{fig:plot1}.

We recall that only the axions that have interacted with the external magnetic field and converted into photons are the ones that are observed in the laboratory frame. Assuming $B_{\rm ext}/{\sqrt{2} M} \ll m$ in (\ref{eqn:cases}), it is easy to show that the number density of observed photons is given by
\begin{equation}
N_L \approx
   \frac{B_{\rm ext}^2}{4 M^2} \dfrac{\pi^{2}\,\xi^{4}}{2^{7}} \dfrac{\nu^{4}}{m^3}~\Phi^{2}\,\mbox{e}^{-\Phi^{2}}. 
\end{equation}

A question which immediately arises is how this mechanism of axion-like particle production compares, for example, with the predicted axion flux from the Sun. We take the coupling coefficient to be $1/M = 2\times 10^{-19}\,\,(m/{\rm eV})$ eV$^{-1}$ \cite{Raffelt1986}.
Consider axion-like particles produced incoherently by $\sim 10^{12}$ oscillating electrons confined in laser
focal spot of diameter $\lambda \sim 100$ $\mu$m, conditions that are achievable in high-power laser experiments. We see the number of axion-like pairs detected per laser shot is given by (assuming $\Phi=2$)
\begin{equation}
\begin{split}
\Delta N_L \approx 300 \left(\frac{N_e}{10^{12}}\right) \left(\frac{\lambda}{100\,\,{\rm \mu m}}\right)^2 \left(\frac{\tau}{100\,\textrm{fs}}\right) \\
\times  \left(\frac{\rm meV}{m}\right) \left(\frac{B_{\rm ext}}{\rm kG}\right)^2 \left(\frac{I_L}{10^{18}\,\,{\rm W/cm^2}}\right)^2,
\end{split}
\end{equation}

where $I_L$ is the laser intensity (in W/cm$^2$).
On the other hand, the number of invisible axions produced every seconds by the Primakoff process in the Sun is given by \cite{Raffelt1986}
\begin{equation}
\Delta N_{\rm Sun} \approx 8.7 \times 10^{42} \left(\frac{m}{\rm eV}\right), 
\end{equation}
which is much larger that $N_L$.
However, suppose axions are emitted isotropically, a detector on Earth of area $A_{d}$ would receive:
\begin{equation}
\Delta N_{\rm helioscope} \approx 3 \times 10^{19} \left(\frac{A_{d}}{\rm m^{2}}\right) \left(\frac{m}{\rm eV}\right).
\end{equation}
Of those, only a tiny fraction will be regenerated into photons (through the $1/M$ coupling), as the axion mass is predicted to lie in the range $1.5\,\textrm{meV} < m < 50\,\mu\textrm{eV}$ \cite{borsanyi16}. Thus, the effective number of measurable invisible axions that the laser-based set-up produces is potentially superior to sun-based searches. Moreover, if the laser repetition rate is significantly higher than a few Hz (as feasible in the foreseeable future), then an axion search of the type proposed here could become competitive against other possible laser-based approaches \cite{Zavattini2006,fouche,Mendonca2007}. 

\bibliography{references}

\end{document}